\documentclass[5p,twocolumn,lineno]{elsarticle}

\usepackage{amssymb,amsmath}
\usepackage{bm}
\usepackage{textcomp}
\usepackage{subfigure}
\usepackage{braket}
\usepackage{float}
\usepackage{xcolor}
\usepackage{cleveref}
\usepackage{multirow}
\usepackage{icomma}
\usepackage{stackrel}

\newcommand{\q}{q}

\begin{document}

\begin{frontmatter}

\title{
From the Boltzmann equation with non-local correlations to a  standard non-linear Fokker-Planck equation}

\author{Airton Deppman}
 \address{Instituto de F\'{\i}sica -  Universidade de São Paulo, São Paulo 05508-090, Brazil 
 }
\ead{deppman@usp.br}

\author{Alireza K.~Golmankhaneh}
 \address{Department of Physics, Urmia Branch, Islamic Azad University, Urmia, Iran 
 }
\ead{alirezakhalili2005@gmail.com}

\author{Eugenio~Meg\'{\i}as}
\address{Departamento de F\'{\i}sica At\'omica, Molecular y Nuclear and Instituto Carlos I de F\'{\i}sica Te\'orica y Computacional, \\
Universidad de Granada, Avenida de Fuente Nueva s/n, 18071 Granada, Spain 
}
\ead{emegias@ugr.es}

\author{Roman~Pasechnik}
\address{Department of Physics, Lund University, 221 00 Lund, Sweden}
\ead{Roman.Pasechnik@hep.lu.se} 

\begin{abstract}
In this work, we study the formal connections between the non-linear Fokker-Planck Equation associated with the non-additive entropy and the Boltzmann Equation with the non-additive correlation functional. The collisional term following the $q$-algebra is adopted.  In the derivation of the non-additive Fokker-Planck Equation, two constraints are imposed on the final result: i) that the entropic index $q$ is a characteristic parameter of the non-additive systems with a value that does not change with time, and ii) that for $q \rightarrow 1$ a smooth transition for the standard Fokker-Planck Equation is obtained.
\end{abstract}

\end{frontmatter}



The Fokker-Planck Equation (FPE) has been extensively used in many branches of Physics, describing the time evolution of the probability density for dynamic systems.  The Non-linear FPE (NLFPE) appears in several phenomena, e.g., anomalous diffusion. Systems governed by the entropy $S_{\q}$~\cite{Tsallis,TsallisBook} have been associated with a particular NLFPE~\cite{PLASTINO1995347}, where non-local effects give rise to a particular form of NLFPE, which has applications in the dynamics of self-gravitating systems~\cite{Shiino,Chavanis}, overdamped systems~\cite{PlastinoWedermannTsallis2022} and super- and sub-diffusive transport regimes~\cite{CasasNobre}. That equation was used to study granular medium~\cite{TsallisBukman}, and the predictions presented in that work were experimentally proven to be correct~\cite{CombeAtman}. In this work, FPE associated with the non-additive systems will be called non-additive FPE or simply qFPE.

In recent years, both FPE and NLFPE have found increasing interest in High Energy Physics (HEP) because it is a tool to investigate the dynamics of heavy quarks in the Quark Gluon Plasma (QGP)~\cite{Svetitsky:1987gq,Walton:1999dy}. The HEP data have evidenced the heavy-tailed momentum distributions that are described by the q-exponential function, and therefore the HEP is associated with the nonadditive entropy, $S_{\q}$. The emergency of the nonextensive statistics in non-Abelian Quantum Field Theory (QFT) has been associated with the thermofractal structures~\cite{Deppman-PRD-2016,DMM-Federico-2018} resulting from the scaling properties of the field theory that are manifested in the renormalization group equation~\cite{DMM-PRD-2020}. As a consequence of the fractal structure, the QCD running-coupling is obtained by an analytical formula. The parameter $q$, that in the nonextensive statistics represents the entropic index, can be calculated in terms of the QCD fundamental parameters, which are the number of colours, $N_c$, and the number of flavours, $N_f$. Hereby, the qFPE might be a useful tool for describing the dynamics of quarks in the QGP.

In the present work, we study formal connections between the qFPE and the standard Boltzmann Equation endowed with a correlation term proposed in Refs.~\cite{Lima:2001lgd, Lavagno:2002hv}. The stationary solutions for the generalized equation are $q$-exponentials and are described by the $S_{\q}$ statistics. The origins of the new correlation term are associated with the finite size of the system~\cite{Megias:2022yrw} and, mathematically, with the $q$-algebra~\cite{Borges-qCalculus}.

The dynamics of complex systems following the non-additi\-ve entropy must necessarily be different from those following the Boltzmann statistics, and the main difference dwells in the collisional term of the dynamical equation. As a consequence, the Fokker-Planck Equation must be modified, but one can expect that both the dynamical equation and the Fokker-Planck Equation recover the correspondent equation for the Boltzmann statistics in the limit of $q \rightarrow 1$.

Starting from the Boltzmann Equation, if $f(\mathbf x, \mathbf p, t)$ is the probability density for a system over the phase-space, the Boltzmann Equation is given by
\begin{equation}
   \frac{\partial f(\mathbf p, t)}{\partial t}= - \nabla_{\mathbf p} f( \mathbf p, t) \cdot \mathbf F + C[f] \,,
\end{equation}
where $C[f]$ is the integral collision term, and
\begin{equation}
    f( \mathbf p, t) = \int d^3x f(\mathbf x, \mathbf p, t) \,.
\end{equation}
Here, it is assumed that the density $f(\mathbf x, \mathbf p, t)$ vanishes for $|\mathbf x| \rightarrow \infty$. In the following, the symbol $t$ in the argument of the density will be omitted for convenience. In the absence of external force, $\mathbf F=0$, and the equation reduces to
\begin{equation}
   \frac{\partial f(\mathbf p)}{\partial t}=  C[f] \,.
\end{equation}

For a quantum system, the collision term is expressed in terms of the transition matrix, $\cal M$, as 
\begin{equation}
 \begin{split}
&    C[f( \mathbf p)]= \frac{1}{2\omega_p}  \int \frac{d^3p'}{2\omega_{p'}} \int \frac{d^3q}{2\omega_{q}} \int \frac{d^3q'}{2\omega_{q'}} |{\cal M} |^2 \times \\
&\; \delta^{(3)}\left((\mathbf p'-\mathbf p)-(\mathbf q-\mathbf q') \right)  \left( h[f(\mathbf p),f(\mathbf q)]- h[f(\mathbf p'),f(\mathbf q')]\right)\,, \label{Qcollision}
 \end{split}
\end{equation}
where, usually, one has
\begin{equation}
     h[f(\mathbf p),f(\mathbf q)]=f( \mathbf p)f( \mathbf q)
     \,. \label{Bcollision}
\end{equation}
At the stationary state, the density will be described by an exponential function. Initially, distribution functions of the form $f \equiv f(\bm \beta \cdot \mathbf p)$, with $\bm \beta=\beta_x \mathbf i + \beta_y \mathbf j + \beta_z \mathbf k$ will be considered.

The correlation term in Eq.~(\ref{Bcollision}) plays a central role in the derivation. Defining the transition rate, $w(\mathbf p, \mathbf k)$ as
\begin{equation}
 w(\mathbf p, \mathbf k)= \frac{1}{2\omega_p}  \frac{1}{2\omega_{p'}} \frac{1}{2\omega_{q'}}\int \frac{d^3q}{2\omega_{\mathbf q}} |{\cal M} |^2 f(\mathbf q)\, \label{transitionrate}
\end{equation}
with $\mathbf k=\mathbf p' - \mathbf p$, it results~\footnote{Notational remark: $(wf)(\mathbf a)=w(\mathbf a)f(\mathbf a)$. Wherever possible, the second argument, $\mathbf k$, will be omitted in $w$ and $(wf)$.}
\begin{equation}
     C[f( \mathbf p)] = \int d^3k \, [(wf)(\mathbf p + \mathbf k)- (wf)(\mathbf p)] \,.
\end{equation}
Expanding the product $(wf)(\mathbf p)$ up to the second order, to satisfy the Pawula theorem~\cite{Pawula}, the usual Fokker-Planck Equation is obtained.

For non-additive systems, the correlation function is given by
\begin{equation}
    h[f(\mathbf p),f(\mathbf q)]_q=f(\mathbf p)\otimes_q f(\mathbf q)\,, \label{qcorrelation}
\end{equation}
where the symbol $\otimes_q$ represents the $q$-product~\cite{Borges-qCalculus}. In Refs.~\cite{Lima:2001lgd,Lavagno:2002hv} the authors show that the $q$-exponential distribution can be obtained at the stationary state if, instead of Eq.~(\ref{Bcollision}), the collision term is expressed in terms of
\begin{equation}
    h[f(\mathbf p),f(\mathbf q)]_q=\left[f( \mathbf p)^{1-q}+f( \mathbf q)^{1-q}-1\right]^{\frac{1}{1-q}} \,. \label{Tcollision}
\end{equation}
The term on the right-hand side of the equation above is the algebraic representation of the $q$-product, and it results in the regular product of functions when $q \rightarrow 1$.

The main aspects related to the non-additive functional in Eq.~(\ref{Tcollision}) were discussed in Ref.~\cite{Tsallis:2022vwi}, and in Ref.~\cite{Megias:2022yrw} it was shown that the origin of the correlation might be related with topological characteristics of the phase-space. Because in the non-additive case the correlation term changes, it is necessary to derive the extended Fokker-Planck Equation for this case. This will be done in this work 

It will be considered that $\mathbf k$ can be made sufficiently small. One can observe, from Eq.~(\ref{qcorrelation}), that $f(\mathbf k) \rightarrow 1$ for $\mathbf k \rightarrow 0$. Using the first-order approximation of the right-hand side of Eq.~(\ref{Tcollision}) it results
\begin{equation}
 f(\mathbf p) \otimes_q f(\mathbf k)=f(\mathbf p) + f(\mathbf p)^q {\bm {\beta}}  \cdot {\mathbf k} \,, \label{linearapprox}
\end{equation}
where $\bm{\beta} \equiv \frac{\partial f}{\partial \mathbf k} \big|_{\mathbf k = 0}$ is the linear coefficient of the first order approximation of the function $f(\mathbf k)$. According to the rules of the $q$-algebra, if $\mathbf k$ is an infinitesimal vector, the $q$-product is\footnote{The $q$-calculus was introduced in Ref.~\cite{Borges-qCalculus}, where the $q$-product is defined in accordance with the r.h.s. of Eq.~(\ref{Tcollision}) in the present paper. In Eq.~(\ref{qprod}), we use the dual derivative operator, defined by Eq.~(20) in Ref.~\cite{Borges-qCalculus}.}
\begin{equation}
 f(\mathbf p) \otimes_q f(\mathbf k)=f(\mathbf p)\left[1+\frac{\bm{\beta} \cdot \mathbf k}{1+ (1-q) \bm{\beta} \cdot \mathbf p}\right]\,. \label{qprod}
\end{equation}
Comparing the last two equations, it results in
\begin{equation}
 f(\mathbf p)^q =\frac{f(\mathbf p)}{1+ (1-q) \bm{\beta} \cdot \mathbf p} \,. \label{derivativeconstraint}
\end{equation}

The group of transformations of thermofractals is isomorphic with the $q$-algebra~\cite{Deppman-Physics-2021}, so its evolution is also described by Eq.~(\ref{derivativeconstraint}). Moreover, due to the fractal structure with scaling invariance, it also follows that
the product $f(\mathbf p)\otimes_q f(\mathbf k)=f(\mathbf p + \mathbf k)$. Here, we assume that this equation is valid in general, for all non-additive systems. From Eq.~(\ref{linearapprox}) and using the result above, it results that
\begin{equation}
  f(\mathbf p+\mathbf k)-f(\mathbf p) =  \frac{\partial f}{\partial \mathbf p}  \cdot  \mathbf k= f(\mathbf p)^q \bm{\beta} \cdot \mathbf k   \,, 
\end{equation}
therefore
\begin{equation}
    \frac{\partial f(\mathbf p)}{\partial \mathbf p} = f(\mathbf p)^q \bm{\beta} \,. \label{qderivative}
\end{equation}

The equation above, if understood in a proper way, is the central point of the formal connection between the Boltzmann Equation and the non-additive Fokker-Planck Equation. Notice that, already in the Boltzmann equation, there is the introduction of the $q$-product in the place of the regular product. From an epistemological point of view, this is the necessary gap for going from the Boltzmann statistics to the non-additive statistics. However, the regular product plays a central role in the derivation of the FPE, so it is necessary to replace the $q$-product by the regular product. In doing so, the epistemological gap must be placed at another point, and Eq.~(\ref{qderivative}) will be used for that. Here, Eq.~(\ref{qderivative}) is to be understood as valid for any function $f$, and so it imposes constraints on the derivative operator that will be incorporated in the qFPE to be derived.

The correlation function in Eq.~(\ref{qcorrelation}) can be written in the form of a regular product if one constrains the distribution $f(\mathbf p)$ to satisfy the constraint on the derivative given by Eq.~(\ref{qderivative}), since if the latter equation is valid, the first one necessarily holds. Then, the collision term of the Boltzmann Equation assumes the same form for the non-additive systems as for the systems following the Boltzmann statistics, and the transition rate, $w(\mathbf p,\mathbf k)$, can be written in the same way as in Eq.~(\ref{transitionrate}). For clarity, the symbol $\bar{\partial}/\partial \mathbf p$ will be used to identify the special derivative operator which satisfies the non-additivity condition in Eq.~(\ref{qderivative}). This symbol will be eliminated in favour of the standard derivative in the process to obtain the qFPE, and its formal structure will be described by the relations~(\ref{specialderivative}), below.

Expanding the product $wf$ up to the second order, the terms depending on $\mathbf p+\mathbf k$ result in
\begin{equation}
 \begin{split}
   (wf)(\mathbf p+\mathbf k)-  (wf)(\mathbf p)= & \sum_i k_i \frac{\bar{\partial} (wf)(\mathbf p)}{\partial p_i} + \\ & 
   \frac{1}{2} \sum_{i,j}  k_i k_j \frac{\bar{\partial}^2}{\partial p_i \partial p_j} (w f)(\mathbf p)  \,.
 \end{split}
\end{equation}

Because only the transition rate depends on $\mathbf k$, it is possible to integrate over $d^3k$, obtaining
\begin{equation}
    \frac{\partial f}{\partial t}- \frac{\bar{\partial}}{\partial p_i} \left[A_i f+ \frac{\bar{\partial} (B_{ij}f) }{\partial p_j}\right]=0 \,, \label{FPE}
\end{equation}
where
\begin{equation}
    A_i=\int d^3k \, w(\mathbf p, \mathbf k) k_i \,,
\end{equation}
and
\begin{equation}
    B_{ij}= \frac{1}{2} \int d^3k \, w(\mathbf p, \mathbf k) k_i k_j \,.
\end{equation}
The summation rule over repeated indexes is adopted in Eq.~(\ref{FPE}).

The result obtained in the last equation is, formally, the usual term for the standard Fokker-Planck Equation, unless by the use of the special derivatives. The objective of the present is to obtain the form of the Fokker-Planck Equation for non-additive systems in terms of the standard derivatives. This task is essentially different from the one carried out in Ref.~\cite{CuradoNobre1}. Here, the non-additive Fokker-Planck Equation will be derived from the properties of the correlation functional. The following constraints are imposed on the qFPE:
\begin{enumerate}
    \item that the fundamental aspects of the non-additive systems can be completely incorporated into the mathematical description of the dynamics by imposing the relations of the derivatives of the distribution presented in Eq.~(\ref{qderivative});
    \item that the parameter $q$ is a well-defined number during the diffusive process, that is, $q$ is constant and unique for a given system;
    \item that the generalized FPE smoothly recover the standard form of the Boltzmann statistics when $q \rightarrow 1$.
\end{enumerate}
It will be shown that the non-linear Fokker-Planck Equation proposed in Ref.~\cite{PLASTINO1995347} is obtained from the Boltzmann Equation by solely assuming the non-additive correlation function and the $q$-algebra.

To incorporate the deformed algebra in terms of the standard derivative operators, note that from Eq.~(\ref{qderivative}) it results that
\begin{equation}
 \frac{\partial f}{\partial p_i}=- \beta_i f^q=- \beta_i f^{q-1}f \,, \label{1stDerivative}
\end{equation}
and hereby
\begin{equation}
   \frac{\partial}{\partial p_j} \left(f^{1-q} \frac{\partial f}{\partial p_i}\right) = - \beta_i \frac{\partial  f}{\partial p_j} \,.
\end{equation}
Considering that
\begin{equation}
    f^{1-q} \frac{\partial f}{\partial p_i}=\frac{1}{(2-q)}\frac{\partial f^{2-q}}{\partial p_i}\,,
\end{equation}
we have
\begin{equation}
     \beta_i \frac{\partial f}{\partial p_j} = - \frac{1}{(2-q)}\frac{\partial^2 f^{2-q}}{\partial p_i \partial p_j}\,. \label{2ndDerivative}
\end{equation}

All constraints on the qFPE can be satisfied by noting that, near the stationary state, the solutions of the qFPE are expected to be of the $q$-exponential form. Then, all terms in the qFPE must return a function that has the same exponent, and from Eqs.~(\ref{1stDerivative}) to~(\ref{2ndDerivative}) it results that one can make the substitutions
\begin{equation}
    \begin{cases}
        {\bar\partial f}/\partial p_i \rightarrow  \partial f/\partial p_i \\
        \bar{\partial}^2 f/(\partial p_i \partial p_j) \rightarrow  \partial^2 f^{2-q}/(\partial p_i \partial p_j) \,. \label{specialderivative}
    \end{cases}
\end{equation}
With those substitutions, the qFPE will be
\begin{equation}
    \frac{\partial f}{\partial t}- \frac{\partial}{\partial p_i} \left[A_i f+ \frac{\partial (B_{ij}f^{2-q})}{\partial p_j}\right]=0 \,. \label{qFPE}
\end{equation}
Observe that the equation above results in the standard FPE when $q=1$, and all terms will contribute with the same value for $q$, because of the result obtained in Eq.~(\ref{2ndDerivative}).

If $\mathbf p \sim 0$ then $A_i \rightarrow \gamma p_i$ and $B_{ij} \rightarrow D \delta_{ij}$, and 
the qFPE reduces to
\begin{equation}
    \frac{\partial f}{\partial t}- \gamma \frac{\partial }{\partial \mathbf p}\cdot (\mathbf p f) - D \frac{\partial^2 f^{2-q}  }{ \partial \mathbf p^2} =0 \,. \label{qFPElinearized}
\end{equation}

It is possible to show that the qFPE above has a stationary solution of the form (for 1-dimensional cases)
\begin{equation}
    f(p)=A \left[ 1-(1-q)\beta (p-p_{M}) \right]^{\frac{1}{1-q}} \,,
\end{equation}
where $p_M$ is the centre of the distribution at the stationary regime.
The qFPE can be generalized by using $\bm \beta=\sum_i \beta_i  \mathbf e_i$, where $\mathbf e_i$ are  unitary vectors. The general solution for the equation will be similar to the one above. 

For more complex cases, a stationary solution can be obtained only if the external force, $\mathbf F$, in the Boltzmann Equation is not null. As an example, if $\mathbf F= \mathbf a + b \mathbf p$, a general solution for the qFPE can be obtained in the form
\begin{equation}
    f(\mathbf p)=A \left[ 1-(1-q) \bm{\beta} \cdot (\mathbf p - \mathbf{p_{M}}(t))^2 \right]^{\frac{1}{1-q}} \,,
\end{equation}
that represents a $q$-Gaussian function centered at $\mathbf{p_{M}}(t)$$ $~\cite{TsallisBukman}. This solution presents an asymptotically stationary state for $t \rightarrow \infty$.

To complete the derivation of the qFPE, it will be shown that Eq.~(\ref{qFPE}) is the only equation that satisfies the constraints imposed initially. One could consider a general NLFPE of the form
\begin{equation}
 \frac{\partial f^a}{\partial t} = \gamma \frac{\partial f^b}{\partial \mathbf p} +D\frac{\partial^2 f^c}{\partial  p^2} \,,
\end{equation}
for $a, b, c > 0$. Considering the second constraint we imposed for the derivation of the qFPE and the properties of the derivative operations described by Eqs.~(\ref{1stDerivative})-(\ref{2ndDerivative}), the constants $a$, $b$ and $c$ must satisfy the relation
\begin{equation}
 \frac{a}{1-q}=\frac{b}{1-q}=\frac{c}{1-q}-1 \,.
\end{equation}
These equalities imply that $a=b$ and $c=a+(1-q)$. For $a=1$, it results

\begin{equation}
   \frac{\partial f}{\partial t}  = \gamma \frac{\partial f}{\partial  p} +D\frac{\partial^2 f^{2-q}}{\partial  p^2} \,.  \label{NAPF2}
   \end{equation}
The equation above is exactly that proposed in Ref.~\cite{PLASTINO1995347}. However, it is always possible to define $q'$ such that
\begin{equation}
 \frac{1}{1-q'}=\frac{a}{1-q}\,,
\end{equation}
and from here one gets that
\begin{equation}
 \frac{c}{1-q}=\frac{2-q'}{1-q'}\,,
\end{equation}
resulting in the same Fokker-Planck generalization as that in Eq.~(\ref{NAPF2}), but in terms of $q'$ instead of $q$, and $f$ satisfying the $q$-algebra with entropic index $q^\prime$. It is important to remind that the scale $\beta$ changes also to $\beta'=\beta a$.

For the case $a, b, c < 0$, one also obtains a similar form of the generalised Fokker-Planck Equation but associated with a different form of the $q$-exponential equation, according to the transformations discussed in Ref.~\cite{Deppman-Physics-2021}. In this case, non-integrable solutions might be found. This result shows that Eq.~(\ref{qFPE}) is the only possible form for the Fokker-Planck Equation that satisfies the constraint defined in the present work.

Concluding, the present work reports the formal connections between the Boltzmann Equation generalized to include systems following the non-additive entropy $S_{\q}$,  and the non-linear Fokker-Planck Equation proposed in Ref.~\cite{PLASTINO1995347}. The approach used here starts by considering the Boltzmann Equation with a collisional term that incorporates  the $q$-algebra in the correlation functional. By considering the properties of the $q$-product and the fact that for a non-additive system the value of the entropic index $q$ represents a well-defined quantity that remains unchanged during the dynamical evolution of that system, the generalized Fokker-Planck Equation is derived. It is shown that the only possible equation satisfying those constraints is the generalized Fokker-Planck Equation given by Eq.~(\ref{qFPE}).

This work evidences some aspects of the qFPE, proposed in Ref.~\cite{PLASTINO1995347}, by clarifying its relations with the Boltzmann Equation. The equation obtained has already been tested experimentally, since the work in Ref.~\cite{CombeAtman} confirms the predictions done in Ref.~\cite{TsallisBukman} by using Eq.~(\ref{qFPE}). The result obtained here is of interest in any application of the dynamical evolution for systems that are non-extensive, such as rock blasting~\cite{RockBlasting}, and fundamental aspects of dynamical systems~\cite{daCosta,Onofrio,Chavanis2},

With the increasing interest in the propagation of quarks in the QGP, the qFPE can be an important tool to model the quark interaction with the medium~\cite{Walton:1999dy,SrivastaPatraKrishna,ParvanTeryaevCleymans}, and multiparticle production~\cite{WilkWlodarkzyk-multiparticle,WongWilkCirtoTsallis,Berrehrah}. The latter application will be considered in future works. In this particular domain, it is fundamental to reckon with the effects of finite quark masses, which leads to a set of coupled equations to tackle the different particle species. This type of phenomenon has been recently addressed, in the context of the non-linear Fokker-Planck Equation, by the work in Ref.~\cite{PlastinoWedemannNobre}.

Both the nonlinear Fokker-Planck equation and the Boltzmann
equation with non-local correlations are subjects of intense research eﬀorts, with applications to various branches of physics, including Astrophysics and High Energy Physics. The study of these two evolution equations and their respective applications has remained, up to now, research areas largely independent of each other. It is, therefore, a remarkable achievement to establish a link between these active lines of inquiry. Although the present work has been motivated by and aimed at applications in high energy physics, the results obtained have a general and fundamental character that makes them relevant also for other areas of physics, particularly in connection with the study of complex systems.

\vspace{0.7cm}
\noindent {\bf Acknowledgements}  
\vspace{0.3cm}

The authors thank Dr Constantino Tsallis and Dr Ernesto Borges for enlightening discussions on the subject of this work.
The authors would like to thank the anonymous reviewer of this paper. The last sentence in the conclusions is taken from his comments on the present work.

E.M. would like to thank the Instituto de F\'{\i}sica and the Instituto de Astronomia e Geof\'{\i}sica of the Universidade de S\~ao Paulo, Brazil, for their hospitality during the final stages of this work. 
A.D. is supported by the Conselho Nacional de Desenvolvimento Cient\'{\i}fico e Tecnol\'ogico (CNPq-Brazil), grant 304244/ 2018-0, by
Project INCT- FNA Proc. No. 464 898/2014-5, and by FAPESP, Brazil grant 2016/17612-7. 
The work of E.M. is supported by the project PID2020-114767GB-I00 funded by MCIN/AEI/10.13039/501100011033, by the FEDER/ Junta de Andaluc\'{\i}a-Consejer\'{\i}a de Econom\'{\i}a y Conocimiento 2014-2020 Operational Program under grant A-FQM178-UGR18, and by Junta de Andaluc\'{\i}a under grant FQM-225. The research of E.M. is also supported by the Ram\'on y Cajal Program of the Spanish MCIN under grant RYC-2016-20678.
R.P.~is supported in part by the Swedish Research Council grants, contract numbers 621-2013-4287 and 2016-05996, as well as by the European Research Council (ERC) under the European Union's Horizon 2020 research and innovation programme (grant agreement No 668679).



\end{document}